\documentclass[a4paper]{article}

\usepackage[english]{babel}
\usepackage[utf8x]{inputenc}
\usepackage[T1]{fontenc}
\usepackage{ upgreek }
\usepackage{gensymb} 
\usepackage{threeparttable}
\usepackage{tabularx}
\usepackage{makecell}
\usepackage{cite}
\usepackage{mathrsfs}
\usepackage{amssymb}
\usepackage{xcolor}
\usepackage{booktabs}
\makeatletter
\newcommand{\ssymbol}[1]{^{\@fnsymbol{#1}}}
\makeatother

\usepackage{subcaption}
\usepackage{amsmath}
\usepackage{graphicx}
\usepackage[colorinlistoftodos]{todonotes}
\usepackage[colorlinks=true, allcolors=blue]{hyperref}

\usepackage{caption}
\usepackage[a4paper,top=2cm,bottom=2cm,left=3cm,right=3cm,marginparwidth=1.75cm]{geometry}

\usepackage{authblk}
\title{ Multi-party Semi-quantum Secret Sharing Protocol based on Measure-flip and Reflect Operations}
\author[1,3]{Jian Li}
\author[2]{Chong-Qiang Ye \thanks{Corredponding author:  chongqiangye@bupt.edu.cn}}
\affil[1]{ School of Information Engineering, Ningxia University, Yinchuan, 750021, China}
\affil[2]{ School of Artificial Intelligence, Beijing University of Posts Telecommunications, Beijing 100876, China.}
\affil[3]{Information Security Center, State Key Laboratory of Networking and Switching Technology,
Beijing University of Posts and Telecommunications, Beijing 100876, China}

\begin{document}
\date{}
  \maketitle

\begin{abstract}
Semi-quantum secret sharing (SQSS) protocols serve as fundamental frameworks in quantum secure multi-party computations, offering the advantage of not requiring all users to possess intricate quantum devices. However, current SQSS protocols mainly cater to bipartite scenarios, with few protocols suitable for multi-party scenarios. Moreover, the multi-party SQSS protocols face limitations such as low qubit efficiency and inability to share deterministic secret information.
 To address this gap, this paper proposes a multi-party SQSS protocol based on multi-particle GHZ states. In this protocol, the quantum user can distribute the predetermined secret information to multiple classical users with limited quantum capabilities, and only through mutual cooperation among all classical users can the correct secret information be reconstructed. By utilizing measure-flip and reflect operations, the transmitted multi-particle GHZ states can all contribute keys, thereby improving the utilization of transmitted particles. Then, security analysis shows that the protocol's resilience against prevalent external and internal threats. Additionally, employing IBM Qiskit, we conduct quantum circuit simulations to validate the protocol's accuracy and feasibility. Finally, compared to similar studies, the proposed protocol has advantages in terms of protocol scalability, qubit efficiency, and shared message types.

\end{abstract}

\section{Introduction}
\label{intro}
Quantum cryptography, whose security is based on quantum laws, contrarily, classical cryptography usually relies on computational complexity assumptions, which is vulnerable to a computationally unbounded adversary. In recent years, many important advances have been made in the field of quantum cryptography\cite{1,2,3,4}.

Quantum secret sharing (QSS)\cite{5} is one of the fundamental applications in quantum cryptography. It aims to allow the sharing of secrets held by a server among multiple participants, with the capability to recover the secret when a sufficient number of participants collaborate. QSS can be regarded as the quantum counterpart of classical secret sharing protocols, designed to enhance protocol security and withstand threats from quantum adversaries. Therefore, QSS has attracted considerable attention from researchers and has been continuously explored and studied. Many significant research findings have been proposed in this field\cite{6,7,8,9,10}. Moreover, due to its ability to satisfy the requirements of multi-user secret sharing, QSS can be applied to scenarios of quantum secure multiparty computation. Serving as a foundational module, QSS can facilitate the construction of secure multiparty computation protocols with various functions, such as quantum sealed-bid auction\cite{11,12,13} and quantum summation\cite{14,15,16,17}.


Although the aforementioned quantum protocols have advantages in terms of security, they all require users in the protocol to have complete quantum capabilities, resulting in significant quantum overhead for the protocol users. However, in the current scenario where quantum resources are insufficient, not all users can access an adequate amount of quantum resources\cite{1}. In reality, most users are unable to afford the burden of complex quantum state preparation and measurement operations. Fortunately, Boyer et al. \cite{18,19} proposed the idea of semi-quantum, which provides a solution for the high cost of quantum resources. In semi-quantum protocols, typically only one user needs to possess full quantum capabilities, while the quantum capabilities of other users are restricted to performing simple operations. These users with restricted capabilities are also referred to as ``classical'' users. They can only carry out computational basis (i.e., $\{|0\rangle,|1\rangle\}$) state preparation and measurements, direct reflect, as well as reordering operations. Semi-quantum protocols reduce the dependency of users on quantum resources, thus attracting widespread attention from researchers. Current research in the field of semi-quantum cryptography includes semi-quantum key distribution\cite{20,21,22,23,24,25}, semi-quantum secure direct communication\cite{26,27,28,29}, semi-quantum secret sharing (SQSS) \cite{30,31,32,33,34,35,36,37,38,39}, semi-quantum secure multi-party computation\cite{40,41,42}, and more. Readers can refer to reference\cite{43} for a detailed overview of semi-quantum cryptography.

The concept of SQSS was first introduced by Li et al.\cite{30} in 2010. In SQSS, the quantum user, Alice, possesses full quantum capabilities and aims to share secret information with ``classical'' users, Bob and Charlie. Only through mutual cooperation between Bob and Charlie can Alice's secret information be recovered. Subsequently, various types of SQSS protocols have been proposed\cite{31,32,33,34,35,36,37,38,39}. For example, in 2015, Xie et al. \cite{31} introduced a novel SQSS protocol where the quantum party can share specific messages with classical parties instead of random messages. In 2023, Xing et al.\cite{32} achieved an SQSS protocol capable of sharing deterministic information using single photons. Later, Hu et al.\cite{33} utilized high-dimensional quantum states as information carriers to design SQSS protocol, extending the protocol to the high-dimensional scenarios. Recently, He et al. \cite{34} pointed out that the protocol based on single particle states proposed by Tian et al.\cite{35} in 2021 that can share certain secret information has security vulnerabilities, and gave an improved method. In 2024, Hou et al.\cite{36} devised an SQSS protocol combining single-particle states and GHZ states with a circular transmission mode. Most of the aforementioned protocols are applicable only to two-party scenarios, meeting the requirement for secret sharing between two ``classical'' users. In practical scenarios, the number of users sharing secret information is often random and more than two. Multi-party SQSS protocols, which are more in line with real-world situations, have become a research hotspot. Currently, several multi-party SQSS protocols have been proposed\cite{37,38,39}, which can share random secret information among multiple ``classical'' users.

However, due to the limitations of ``classical'' users' capabilities in semi-quantum protocols, multi-party SQSS protocols exhibit significant efficiency disadvantages. In more detail, in previous SQSS protocols, only particles measured by ``classical'' users could serve as the basis for establishing mutual keys, while directly reflected particles could only be utilized for eavesdropping. Moreover, to ensure the security of shared secret information, the operations performed by ``classical'' users on received particles were random. Only when certain conditions were met among users (such as all users measuring the particles in computational basis) , it could be used to construct the final secret information. As the number of users increases, satisfying these conditions becomes more stringent. For example, if there are $n$ users, the probability of this situation occurring, where they all choose to measure operations, will be $1/2^n$. This leads to the need for more quantum states, resulting in a significant decrease in protocol efficiency. The low qubit efficiency is the main reason to limit the development of multi-party semi-quantum protocols. Therefore, effectively using the transmitted particles as the secret keys becomes particularly important.

Currently, multi-party SQSS protocols\cite{37,38,39} also grapple with the challenge of low protocol efficiency, as well as the fact that the shared keys are random rather than deterministic. These issues pose difficulties in meeting the practical demand for secret sharing in real-world scenarios.

To address these issues, this work proposes a novel  multi-party semi-quantum secret sharing protocol (MSQSS) based on $(n+1)$-qubit GHZ states. Using measure-flip and reflect operations, "classical" users can encode secret information onto quantum carriers by altering or maintaining the state of the receiving particles. In the proposed protocol, the transmitted multi-particle GHZ states can all contribute to the key, thereby enhancing the utilization efficiency of the transmitted particles.
In addition to GHZ states, the protocol also requires some decoy states (i.e., $\{|0\rangle, |1\rangle, |-\rangle, |+\rangle\}$) to be randomly inserted into GHZ state particles as decoy states to ensure the security of transmitted particles. Security analysis shows that the protocol can withstand typical external and internal attacks. Furthermore, to verify the feasibility of the protocol, we also conducted circuit simulations. Compared to previous protocols, our protocol has advantages in terms of qubit efficiency, protocol scalability, and types of shared messages.

The remaining organization is shown below. Sect. 2 introduces the basic settings of the protocol and properties of $(n+1)$-qubit GHZ states. Sect. 3 provides a detailed illustration of the proposed protocol. Subsequently, in Sect. 4, we analyze the security of the proposed protocol. In Sect. 5, we conduct circuit simulation and discussion. Finally, in Sect. 6 we perform a summary of the protocol.

\section{Preliminary knowledge}
In this part, we will introduce the basic requirements of the protocol and the quantum resources used by the protocol. 

The first is an introduction to the basic settings of the proposed MSQSS protocol. In our protocol, quantum user, Alice, wants to share specific messages $K(k_1,k_2,\dots,k_L)$ with $n$ classical Bob$_i (i=1,2,\dots,n)$. For the secret information $K$ shared by Alice, the correct secret information can only be recovered when all ``classical'' users cooperate with each other. No single user can independently recover $K$. Here, Bob$_i$ is limited to the following operations: 
\begin{itemize}
\item[(1)] \textbf{Measure-flip}: measure the qubit in basis $\{|0\rangle,|1\rangle\}$ and regenerate one in the opposite state (e.g., $|0\rangle \rightarrow |1\rangle$, $|1\rangle \rightarrow |0\rangle$).

\item[(2)] \textbf{Reflect}: reflect the qubit directly.

\item[(3)] \textbf{Reorder}: reorder the qubit via delay lines.

\end{itemize}

Then we introduce the quantum resources used in the protocol.
In our protocol, $(n+1)$-qubit GHZ states play a crucial role as key transmission particles, facilitating the interaction of information between quantum user and classical users. According to the description in reference\cite{44}, it can be represented as:
\begin{equation}
\begin{split}
|G^{\pm}_{k}\rangle   = 
		\frac{1}{\sqrt 2} (\left|0a_1a_2\dots a_n\right\rangle \pm\left|1\bar{a}_1\bar{a}_2\dots\bar{a}_n\right\rangle ),
\end{split}
\end{equation}	
where $k\in \{0,1,\dots,2^{n}-1\}$, and $0a_1a_2\dots a_n$ is the binary representation of $k$ in an $(n+1)$-bit string, i.e., $k=\sum^{n}_{i=1}a_i\times2^{n-i} $. Note that $\bar{a}_i$ and $a_i$ satisfy the relationship of $\bar{a}_i\oplus a_i =1$, where $i=1,2,\dots,n$. For a single-qubit measurement on the $(n+1)$-qubit GHZ state, we will randomly get $0a_1a_2\dots a_n$ or $1\bar{a}_1\bar{a}_2\dots\bar{a}_n$. Then the value of $k$ can be calculated based on the measurement results. Specifically, if the measurement result of the first qubit is 0, the measurement results of the remaining qubits are equivalent to the binary expansion of $k$. If the measurement result of the first qubit is 1, then the measurement results of the remaining qubits, after being flipped, are equivalent to the binary sequence of $k$.

Furthermore, for states $|G^{+}_{k}\rangle$ and $|G^{+}_{k^{\prime}}\rangle=\frac{1}{\sqrt 2} (\left|0a^{\prime}_1a^{\prime}_2\dots a^{\prime}_n\right\rangle +\left|1\bar{a}^{\prime}_1\bar{a}^{\prime}_2\dots\bar{a}^{\prime}_n\right\rangle )$, we have
\begin{equation}
	\langle G^{+}_{k}|G^{+}_{k^{\prime}}\rangle =
	\delta_{k,k^\prime}, 
\end{equation}
where $\delta_{k,k^\prime}$is an impulse function, and $k  = \sum^{n}_{i=1}a_i\times2^{n-i} $, $k^\prime =\sum^{n}_{i=1}a^{\prime}_i\times2^{n-i}$.
Only when $k$ equals $k^\prime$, are states $|G^{+}_{k}\rangle$ and $|G^{+}_{k^{\prime}}\rangle$ the same. If any qubit changes, it will cause $k \neq k^\prime$, thus it can be further concluded that states $|G^{+}_{k}\rangle$ and $|G^{+}_{k^{\prime}}\rangle$ are different. 

Let $r_i=a_i\oplus a^{\prime}_i$, then the values of $k$ and $k^\prime$ can be connected through $r_i$. As long as any two of them are known, the other can be derived. For example, if $k^\prime$ (i.e.,$a^{\prime}_1 a^{\prime}_2\dots a^{\prime}_n$) and $r_i$ are known, then $k$ can be derived as follows
\begin{equation}
\begin{aligned}
k&=(a^{\prime}_1\oplus r_1)\times2^{n-1}+(a^{\prime}_2\oplus r_2)\times2^{n-2}+\dots+(a^{\prime}_n\oplus r_n)\times2^{0}\\
&=a_1\times2^{n-1}+a_2\times2^{n-2}+\dots+a_n\times2^{0}
\end{aligned}.
\end{equation}

The protocol settings and information carriers have been introduced above. Now let's delve into the core idea of the protocol. In this paper, ``classical'' users encode information by altering the state of the receiving particles or keeping it unchanged through the utilization of \textbf{Measure-flip} and \textbf{Reflect} operations. Furthermore, for a multi-particle GHZ state $|G^{+}_{k}\rangle$, any change in the qubits will result in a change in the value of $k$. Thus, ``classical'' users can establish a relationship between the changes in $k$ and the two different operations employed. By analyzing the operation information and the corresponding changes in $k$ values after the quantum state changes, the initial $k$ value, representing the secret information to be shared by the quantum user, can be deduced. Unlike previous protocols that rely on measurement outcomes as keys, this protocol establishes keys among users by utilizing operations performed by ``classical'' users.


\section{ The proposed MSQSS protocol}

In this section, we will give the specific steps of the protocol and analyze the correctness of the protocol.
\subsection{Protocol description}

\textbf{Step 1:} According to the secret $K(k_1,k_2,\dots,k_L)$, Alice prepares $L$ $(n+1)$-qubit GHZ states, and the $j$-th GHZ state is
\begin{equation}
\label{initial state}
|G^{+}_{k_{j}}\rangle= |G(g^{j}_{0},g^{j}_{1},\dots,g^{j}_{n})\rangle  = 
		\frac{1}{\sqrt 2} \left(|0a^{j}_{1}a^{j}_{2}\dots a^{j}_{n}\rangle +|1\bar{a}^{j}_{1}\bar{a}^{j}_{2}\dots\bar{a}^{j}_{n}\rangle \right),
\end{equation}			 			
where $g^{j}_{0},g^{j}_{1},\dots,g^{j}_{n}$ represent the qubits in $|G^{+}_{k_{j}}\rangle$, $j\in (1,2,\dots,L)$ and $k_j=\sum^{n}_{i=1}a^{j}_{i}\times 2^{n-i}$. Then, Alice picks out all the qubits to construct $n+1$ sequences $S_0,S_1,\dots,S_n$, which can be denoted as follows.
\begin{equation}
\begin{split}
\label{sequence}
&S_0:  g^{1}_{0},g^{2}_{0},\dots,g^{L}_{0},\\
&  \qquad \vdots\\
&S_i:  g^{1}_{i},g^{2}_{i},\dots,g^{L}_{i},\\
&  \qquad \vdots\\
&S_n:  g^{1}_{n},g^{2}_{n},\dots,g^{L}_{n}.
\end{split}
\end{equation}	
 			 			
\textbf {Step 2:} Alice generates $n$ groups of decoy photons $D_1,D_2,\dots,D_n$, where each group has $L$ particles, and each is randomly taken  from the set $\{|0\rangle, |1\rangle, |+\rangle,|-\rangle \}$. After that, $D_i$ is inserted by Alice into $S_i$ to produce a new sequence $S^{*}_{i}$. Subsequently, Alice keeps $S_0$ in her hands and sends $S^{*}_1,S^{*}_2,\dots,S^{*}_n$ to Bob$_1$, Bob$_2$,$\dots$, Bob$_n$, respectively.

\textbf {Step 3:} For each received qubit, Bob$_i$ first randomly chooses the \textbf{Measure-flip} or \textbf{Reflect} operations. After that, he performs \textbf{Reorder} operations on these qubits and sends them back to Alice. For convenience, the rearranged sequence is labled as $S^{\prime}_i$. Note that, in this step, Bob$_i$ creates a $2L$ bits string $R_{i} (r^{1}_{i},r^{2}_{i},\dots,r^{2L}_{i})$ to record his operations. If he chooses the \textbf{Reflect} operation on the $h$-th qubit, he sets $r^{h}_{i}=0$. Otherwise, he sets $r^{h}_{i}=1$. Here, $h=1,2,\dots, 2L$. 

\textbf {Step 4:} After Alice stores all the received qubits, Bob$_i$ announces the order of sequence $S^{\prime}_i$. Then, Alice recovers the original order of the sequence and picks out all the decoy photons. Alice informs Bob$_i$ to announce his operations on the decoy photons. According to Bob$_i$'s operations, Alice measures the decoy photons and discusses the correctness of the measurement results with Bob$_i$. Table \ref{tab1} lists all 8 possible situations. For example, when Bob$_i$ chooses the \textbf{Reflect} operation, Alice's measurement results should be identical to her initial states. If the error rate for the above cases exceeds the threshold, the protocol will be terminated and restarted.
\begin{table}
\centering
\caption{Eavesdropping check of the decoy photons}
\label{tab1}     
\centering
\begin{threeparttable}
\begin{tabular}{p{0.1\linewidth}<{\centering}p{0.14\linewidth}<{\centering}p{0.15\linewidth}<{\centering}p{0.15\linewidth}<{\centering}p{0.17\linewidth}<{\centering}} 
 \Xhline{1.0pt}\noalign{\smallskip}
 Initial state & Bob's operations & Measurement results of Bob & Alice's operations  & Measurement results of Alice \\
\hline\noalign{\smallskip}
 $\left|0\right\rangle$ & Measure-flip & $\left|0\right\rangle$ & $M_Z$ & $\left|1\right\rangle$ \\
 \noalign{\smallskip}
 $\left|0\right\rangle$ & Reflect      & / & $M_Z$ & $\left|0\right\rangle$ \\
 \noalign{\smallskip}
 $\left|1\right\rangle$ & Measure-flip & $\left|1\right\rangle$ & $M_Z$ & $\left|0\right\rangle$ \\
 \noalign{\smallskip}
 $\left|1\right\rangle$ & Reflect      & / & $M_Z$ & $\left|1\right\rangle$\\ 
 \noalign{\smallskip}
 $\left|+\right\rangle$ & Measure-flip & $\left|0\right\rangle$ or $\left|1\right\rangle$ & $M_Z$ & $\left|1\right\rangle$ or $\left|0\right\rangle $ \\
 \noalign{\smallskip}
 $\left|+\right\rangle$ & Reflect      & / & $M_X$ & $\left|+\right\rangle$ \\
 \noalign{\smallskip}
 $\left|-\right\rangle$ & Measure-flip & $\left|0\right\rangle$ or $\left|1\right\rangle$ & $M_Z$ & $\left|1\right\rangle$ or $\left|0\right\rangle$ \\
 \noalign{\smallskip}
 $\left|-\right\rangle$ & Reflect      & / & $M_X$ & $\left|-\right\rangle$\\ 
\noalign{\smallskip}
\Xhline{1.0pt}
\end{tabular}
\begin{tablenotes}
\footnotesize
\item[] $M_Z$: Measure the qubits with basis $\{|0\rangle,|1\rangle\}$; \quad $M_X$: Measure the qubits with basis $\{|+\rangle,|-\rangle\}$.
\end{tablenotes}
\end{threeparttable}
\end{table}

\textbf {Step 5:} Alice (Bob$_i$) discards the qubits (bits) used to eavesdropping check. After removing decoy photons, the sequence $S^{\prime}_{i}$ is denoted as $S^{\prime\prime}_{i}$. Each Bob$_i$ uses the remaining bits of $R_i$ as his secret, which is labeled as $R^{\prime}_i (r^{\prime 1}_{i},r^{\prime 2}_{i},\dots,r^{\prime L}_{i})$. Then Alice performs single-particle measurement on the qubits from $S_0$ and $S^{\prime\prime}_{1},S^{\prime\prime}_{2},\dots,S^{\prime\prime}_{n}$. The measurement results are denoted as follows.
\begin{equation}
\label{Single-particle Measurement result }
\begin{split}
&M_0 ( m^{1}_{0},m^{2}_{0},\dots,m^{L}_{0}),\\
&  \qquad \vdots\\
&M_i ( m^{1}_{i},m^{2}_{i},\dots,m^{L}_{i}),\\
&  \qquad \vdots\\
&M_n (m^{1}_{n},m^{2}_{n},\dots,m^{L}_{n}),
\end{split}
\end{equation}	
where $m^{j}_{i}$ is the measurement result of qubit $g^{j}_{i}$ after Bob$_i$'s operation. Note that $m^{1}_{0},m^{2}_{0},\dots,m^{L}_{0}$ are the  measurement results of qubits $g^{1}_{0},g^{2}_{0},\dots,g^{L}_{0}$. According to the value of $m^{j}_{0}$, Alice decides whether to flip the values of $m^{j}_{1},m^{j}_{2},\dots,m^{j}_{n}$ or not. After Alice's operations, the values of $m^{j}_{1},m^{j}_{2},\dots,m^{j}_{n}$ are denoted as $m^{\prime j}_{1},m^{\prime j}_{2},\dots,m^{\prime j}_{n}$ (in fact, $m^{\prime j}_{i}$is equal to $a^{j}_{i}\oplus r^{\prime j}_{i}$, which will be shown in Sect. 3.2),  
\begin{equation}
\label{E10}
m^{\prime j}_{i}=m^{j}_{i}\oplus m^{j}_{0}=
\left\{
   \begin{aligned}
   & m^{j}_{i}\oplus 0, \quad   if \quad m^{j}_{0}=0 \\
   & m^{j}_{i}\oplus 1, \quad    if \quad m^{j}_{0}=1 \\
   \end{aligned}
  \right. . 
\end{equation}	
 After that, Alice publishes $m^{\prime j}_{i}$ to Bob$_1$, Bob$_2$,$\dots$, Bob$_n$.

\textbf {Step 6:} According to the values of $m^{\prime j}_{i}$ and $r^{\prime j}_{i}$, 
Bob$_1$, Bob$_2$,$\dots$, Bob$_n$ work together to calculate $k_{1},k_{2},\dots,k_{L}$
 \begin{equation}
\begin{split}
&k_{1}= \sum^{n}_{i=1}(m^{\prime 1}_{i}\oplus r^{\prime 1}_{i})\times2^{n-i}=\sum^{n}_{i=1}a^{1}_{i}\times 2^{n-i},\\
&k_{2}= \sum^{n}_{i=1}(m^{\prime 2}_{i}\oplus r^{\prime 2}_{i})\times2^{n-i}=\sum^{n}_{i=1}a^{2}_{i}\times 2^{n-i},\\
&  \qquad \vdots\\
&k_{L}= \sum^{n}_{i=1}(m^{\prime L}_{i}\oplus r^{\prime L}_{i})\times2^{n-i}=\sum^{n}_{i=1}a^{L}_{i}\times 2^{n-i}.
\end{split}
\end{equation}
Note that only when Bob$_1$, Bob$_2$,$\dots$, Bob$_n$ cooperate can they obtain Alice's secret $K(k_1,k_2,\dots,k_L)$.

\subsection {Correctness of protocol}
Without considering the decoy photons and eavesdropping check, we take the state $|G^{+}_{k_{j}}\rangle$ as an example to prove the correctness of our protocol. 

Firstly, suppose Alice keeps $|0\rangle$ (i.e., $m^{j}_{0}=0$) in her hand and sends $|a^j_1\rangle,|a^j_2\rangle,\dots, |a^j_n\rangle$ to Bob$_1$, Bob$_2$, $\dots$, Bob$_n$, respectively. For each received qubit, Bob$_i$ uses $r^{\prime j}_{i}$ to record his operation. If his operation is \textbf{Measure-flip}, $r^{\prime j}_{i}=1$; if not, $r^{\prime j}_{i}=0$. Then, the sates of $|a^j_1\rangle,|a^j_2\rangle,\dots, |a^j_n\rangle$ are converted to $|a^j_1 \oplus r^{\prime j}_{1}\rangle,|a^j_2\oplus r^{\prime j}_{2}\rangle,\dots, |a^j_n\oplus r^{\prime j}_{n}\rangle$. After that, Alice measures the qubits sent by Bob$_i$, and the measurement results are labeled as $m^{j}_{1}, m^{j}_{2},\dots, m^{j}_{n}$, where $m^{j}_{i}=a^j_i\oplus r^{\prime j}_{i}$. According to the Eq. (7) and $m^{j}_{0}=0$, it holds that 
\begin{equation}
m^{\prime j}_{i}=m^{j}_{i}\oplus 0= a^j_i\oplus r^{\prime j}_{i}.
\end{equation}

Then, we consider the case that Alice keeps $|1\rangle$ (i.e., $m^{j}_{0}=1$) in her hand and sends $|\bar{a}^j_1\rangle,|\bar{a}^j_2\rangle,\dots, |\bar{a}^j_n\rangle$ to Bob$_1$, Bob$_2$, $\dots$, Bob$_n$, respectively. After Bob$_i$'s operations (i.e.,$r^{\prime j}_{i}$), the states of qubits are converted to $|\bar{a}^j_1 \oplus r^{\prime j}_{1}\rangle,|\bar{a}^j_2\oplus r^{\prime j}_{2}\rangle,\dots, |\bar{a}^j_n\oplus r^{\prime j}_{n}\rangle$. After that, Alice measures the qubits sent by Bob$_i$. The corresponding measurement results are denoted as $m^{j}_{1}, m^{j}_{2},\dots, m^{j}_{n}$, where $m^{j}_{i}=\bar{a}^j_i\oplus r^{\prime j}_{i}=1\oplus{a}^j_i\oplus r^{\prime j}_{i}$. According to the Eq. (7) and $m^{j}_{0}=1$, it holds that 
\begin{equation}
m^{\prime j}_{i}=m^{j}_{i}\oplus 1= 1\oplus a^j_i\oplus r^{\prime j}_{i}\oplus 1= a^j_i\oplus r^{\prime j}_{i}.
\end{equation}
Thus, based on the Eqs. (9-10), it holds that $m^{\prime j}_{i}=a^j_i\oplus r^{\prime j}_{i}$.  According to the steps of the protocol, after obtaining $m^{\prime j}_{i}$, Alice will declare it to Bob$_1$, Bob$_2$,$\dots$, Bob$_n$ via a classical channel. Utilizing the  values of $m^{\prime j}_{i}$ and $r^{\prime j}_{i}$, Alice's secret message $k_j$ can be recovered by calculating 
\begin{equation}
\begin{aligned}
k_j&=(m^{\prime j}_{1}\oplus r^{\prime j}_{1})\times2^{n-1}+(m^{\prime j}_{2}\oplus r^{\prime j}_{2})\times2^{n-2}+\dots+(m^{\prime j}_{n}\oplus r^{\prime j}_{n})\times2^{0}\\
&=(a^{j}_{1}\oplus r^{\prime j}_{1}\oplus r^{\prime j}_{1})\times2^{n-1}+(a^{j}_{2}\oplus r^{\prime j}_{2}\oplus r^{\prime j}_{2})\times2^{n-2}+\dots+(a^{j}_{n}\oplus r^{\prime j}_{n}\oplus r^{\prime j}_{n})\times2^{0}\\
&= a^{j}_{1}\times2^{n-1}+a^{j}_{2}\times2^{n-2}+\dots+a^{j}_{n}\times2^{0}
\end{aligned}.
\end{equation}
It can be clearly seen from Eq. (11) that ``classical'' users can cooperate to recover the secret information $k_j$ shared by Alice by using their respective secret information $r^{\prime j}_{i}$ and the information $m^{\prime j}_{i}$ published by Alice.

Therefore, the output of our protocol is correct. By using this protocol, Alice can share specific messages to the ``classical'' parties (Bob$_i$, $i=1,2,\dots,n$), and only the ``classical'' parties cooperate they can recover Alice's messages.

\section{Security analysis}
Decoy photon technology, derived from the BB84 protocol, would detect most eavesdropping\cite{44,45}. Thus, most typical external and internal attacks are invalid to our protocol. The detailed analysis is shown as follows.
\subsection{The intercept-resend attack}
The intercept-resend attack is a typical eavesdropping method. The attacker will intercept the transmitted particles and resends the particles prepared by herself to the target user. Suppose an external eavesdropper, Eve, uses the intercept-resend attack to obtain some useful information. Without loss of generality, consider Eve attacking particles transmitted between Alice and Bob$_i$. Eve may first intercept all the qubits sent from Alice to Bob$_i$ in Step 2. Then, she generates fake qubits in basis $ \{|0\rangle, |1\rangle \}$, and sends these fake qubits to Bob$_i$. After Bob$_i$ performing the operations on these qubits, Eve catches and measures the qubits sent by Bob$_i$ to obtain his operations. However, Eve's attack is invalid since the order of the qubits sent from Bob$_i$ to Alice is entirely secret for Eve. That is, Eve cannot distinguish which qubit is measure-flipped and which is reflected. More seriously, Eve's fake qubits are not identical to Alice's decoy photons, which will make Eve's attack discovered by Alice during the eavesdropping checking in Step 4.
\subsection{The measure-resend attack}
Eve may launch the measure-resend attack to steal some useful information. She first intercepts all the qubits sent by Alice, and measures them in basis$ \{|0\rangle, |1\rangle \}$. Then she sends the measured qubits to Bob$_i$ directly. After that, Bob$_i$ performs the \textbf{Measure-flip} operation or the \textbf{Reflect} operation on the received qubits. Subsequently, Eve intercepts the qubits sent from Bob$_i$ to Alice, and measures these qubits to extract some useful information. However, Alice will detect Eve's attack in the eavesdropping checking. Specifically, if Bob$_i$ chooses the \textbf{Measure-flip} operation, Eve's attack induces no errors. If Bob$_i$ chooses the \textbf{Reflect} operation, Alice can easily detect Eve's attack because Eve's attack destroys the states of $|+\rangle$ and $|-\rangle$. Therefore, adopting the measure-resend attack, Eve will unavoidably be found by Alice.

\subsection{The entangle-measure attack}
The entangle-measure attack attack is one of the typical attack methods. Specifically, Eve will perform $(U_E,U_F)$ on the target qubit and her probe $|0\rangle_E$ and then she will measure her probe to extract useful information. Here, $U_E$ is the attack operator applied on the qubits sent from Alice to Bob$_i$ while $U_F$ is the attack operator applied on the qubits sent from Bob$_i$ to Alice. The effect of $(U_E, U_F)$ on the target qubit is described as follows \cite{20}:
\begin{equation}
\begin{aligned}
U_E|0,0\rangle_{TE} = \alpha |0,e_0\rangle +\beta |1,e_1\rangle,\\
U_E|1,0\rangle_{TE} = \beta |0,e_2\rangle + \alpha |1,e_3\rangle,
\end{aligned}
\end{equation}
and 
\begin{equation}
\begin{aligned}
U_F|0,e_{\omega}\rangle_{TE} = \mu_{\omega} |0,e^{0}_{0,\omega}\rangle +\nu_{\omega} |1,e^{1}_{0,\omega}\rangle,\\
U_F|1,e_{\omega}\rangle_{TE} = \nu_{\omega} |0,e^{0}_{1,\omega}\rangle +\mu_{\omega} |1,e^{1}_{1,\omega}\rangle,
\end{aligned}
\end{equation}
where $T$ and $E$ represent the target qubit and Eve's probe, respectively. Since $U_E$ and $U_F$ are unitary, which imply $|\alpha|^2+|\beta|^2=1$, $|\mu_{\omega}|^2+|\nu_{\omega}|^2=1$, and $\omega=0,1,2,3$.

Firstly, we analyze the effect of $(U_E, U_F)$ on the decoy photons. The effect of $U_E$ on the decoy photons sent by Alice is shown as follows.
\begin{equation}
\begin{aligned}
\label{E17}
U_E|0,0\rangle_{TE} &= \alpha |0,e_0\rangle +\beta |1,e_1\rangle,\\
U_E|1,0\rangle_{TE} &= \beta |0,e_2\rangle + \alpha |1,e_3\rangle,\\
U_E|+,0\rangle_{TE} &= \frac{1}{\sqrt 2}\big(\alpha |0,e_0\rangle +\beta |1,e_1\rangle+\beta |0,e_2\rangle + \alpha |1,e_3\rangle \big),\\
U_E|-,0\rangle_{TE} &= \frac{1}{\sqrt 2}\big(\alpha |0,e_0\rangle +\beta |1,e_1\rangle -\beta |0,e_2\rangle - \alpha |1,e_3\rangle \big).
\end{aligned}
\end{equation}
In Step 4, Alice and Bob$_i$ do the eavesdropping checking. When Bob$_i$ chooses the \textbf{Measure-flip} operation on the qubits $|0\rangle$ and $|1\rangle$, his measurement results should be identical to Alice's initial states. If Eve wants to be undetected, $U_E$ needs to satisfy the conditions: $\beta=0$ and $\alpha=1$. Thus, the Eq. (14) can be reduced to 
\begin{equation}
\begin{aligned}
\label{E18}
U_E|0,0\rangle_{TE} &=  |0,e_0\rangle, \\
U_E|1,0\rangle_{TE} &=  |1,e_3\rangle,\\
U_E|+,0\rangle_{TE} &= \frac{1}{\sqrt 2}\big(|0,e_0\rangle +|1,e_3\rangle \big),\\
U_E|-,0\rangle_{TE} &= \frac{1}{\sqrt 2}\big(|0,e_0\rangle - |1,e_3\rangle \big).
\end{aligned}
\end{equation}

Next, Eve performs $U_F$ on the decoy photons sent by Bob$_i$. According to the Ref.\cite{30}, Eve performing $U_F$ depends on the knowledge acquired by $U_E$. Due to the \textbf{Reorder} operation of Bob$_i$, the order of decoy photons sent from Alice to Bob$_i$ is different from that Bob$_i$ sent to Alice. Hence, there are two cases need to be analyzed.

Case 1: After the \textbf{Reorder} operation, the order of some decoy photons is not changed. According to the Eq. (15), Eve's probe $|0\rangle_E$ is changed to $|e_{0}\rangle$ or $|e_{3}\rangle$. When Bob$_i$ chooses the \textbf{Reflect} operation, the effect of $U_F$ is shown as follows,
\begin{equation}
\begin{aligned}
U_F|0,e_0\rangle_{TE} &= \mu_{0} |0,e^{0}_{0,0}\rangle +\nu_{0} |1,e^{1}_{0,0}\rangle,\\
U_F|1,e_3\rangle_{TE} &= \nu_{3} |0,e^{0}_{1,3}\rangle +\mu_{3} |1,e^{1}_{1,3}\rangle,\\
U_F\left[\frac{1}{\sqrt 2}\big(|0,e_0\rangle_{TE} + |1,e_3\rangle_{TE}\big)\right] 
&=
 \frac{1}{2}|+\rangle\big(\mu_0|e^{0}_{0,0}\rangle+\nu_0|e^{1}_{0,0}\rangle+\nu_3|e^{0}_{1,3}\rangle+\mu_3|e^{1}_{1,3}\rangle\big) \\
&+\frac{1}{2}|-\rangle\big(\mu_0|e^{0}_{0,0}\rangle-\nu_0|e^{1}_{0,0}\rangle+\nu_3|e^{0}_{1,3}\rangle-\mu_3|e^{1}_{1,3}\rangle\big), 
\\
U_F\left[\frac{1}{\sqrt 2}\big(|0,e_0\rangle_{TE} - |1,e_3\rangle_{TE}\big)\right] 
&=
 \frac{1}{2}|+\rangle\big(\mu_0|e^{0}_{0,0}\rangle+\nu_0|e^{1}_{0,0}\rangle-\nu_3|e^{0}_{1,3}\rangle-\mu_3|e^{1}_{1,3}\rangle\big) \\
&+\frac{1}{2}|-\rangle\big(\mu_0|e^{0}_{0,0}\rangle-\nu_0|e^{1}_{0,0}\rangle-\nu_3|e^{0}_{1,3}\rangle+\mu_3|e^{1}_{1,3}\rangle\big).
\end{aligned}
\end{equation}
When Bob$_i$ chooses the \textbf{Measure-flip} operation, the effect of $U_F$ is (Here, we focus on the situation that qubits $|0\rangle$ and $|1\rangle$ are flipped to $|1\rangle$ and $|0\rangle$.)
\begin{equation}
\begin{aligned}
U_F|1,e_0\rangle_{TE} &= \nu_{0} |0,e^{0}_{1,0}\rangle +\mu_{0} |1,e^{1}_{1,0}\rangle,\\
U_F|0,e_3\rangle_{TE} &= \mu_{3} |0,e^{0}_{0,3}\rangle +\nu_{3} |1,e^{1}_{0,3}\rangle.
\end{aligned}
\end{equation} 
If Eve wants to be undetected by Alice, then after the $U_F$ attack, the states of the particles received by Alice should be opposite to their initial states. That is, $U_F$ must satisfy the conditions:
\begin{equation}
\nu_{0}=\nu_{3}=0,  \quad \mu_{0}=\mu_{3}=1, \quad |e^{0}_{0,0}\rangle=|e^{1}_{1,3}\rangle.
\end{equation}

Case 2: After the \textbf{Reorder} operation, the order of some decoy photons is changed. Here, we focus on the decoy photons reflected by Bob$_i$. Suppose Alice sends qubits $|0\rangle$ and $|+\rangle$ to Bob$_i$, and Bob$_i$ sends qubits $|+\rangle$ and $|0\rangle$ to Alice. Based on the Eq. (15), performing $U_E$ on the qubits $|0\rangle$ and $|+\rangle$, the states become to  
\begin{equation}
\begin{aligned}
U_E|0,0\rangle_{TE}U_E|+,0\rangle_{TE}  &= \frac{1}{\sqrt 2} |0,e_0\rangle \big(|0,e_0\rangle +|1,e_3\rangle \big).
\end{aligned}
\end{equation}
For the received qubits, Bob$_i$ performs \textbf{Reflect} operation and swaps the positions of these two qubits. Then the states change to $\frac{1}{\sqrt 2} (|0,e_0\rangle |0,e_0\rangle +|1,e_0\rangle |0,e_3\rangle)$. After performing $U_F$, the states become 
\begin{equation}
\begin{aligned}
\frac{1}{\sqrt 2}\big[ (\mu_{0}|0,e^{0}_{0,0}\rangle+\nu_0|1,e^{1}_{0,0}\rangle)(\mu_{0}|0,e^{0}_{0,0}\rangle+\nu_{0}|1,e^{1}_{0,0}\rangle)\\
+ (\nu_{0}|0,e^{0}_{1,0}\rangle+\mu_{0}|1,e^{1}_{1,0}\rangle)(\mu_{3}|0,e^{0}_{0,3}\rangle+\nu_{3}|1,e^{1}_{0,3}\rangle)\big]
\end{aligned}.
\end{equation}
Substituting equation (18) into equation (20), we can derive
\begin{equation}
\begin{aligned}
&\frac{1}{\sqrt 2}( \mu_{0}|0,e^{0}_{0,0}\rangle \mu_{0}|0,e^{0}_{0,0}\rangle+\mu_{0}|1,e^{1}_{1,0}\rangle \mu_{3}|0,e^{0}_{0,3}\rangle)\\
&=\frac{1}{2}(|+\rangle+|-\rangle)|e^{0}_{0,0}\rangle|0,e^{0}_{0,0}\rangle+\frac{1}{2}(|+\rangle-|-\rangle)|e^{1}_{1,0}\rangle|0,e^{0}_{0,3}\rangle\\
&=\frac{1}{2}|+,0\rangle(|e^{0}_{0,0}\rangle|e^{0}_{0,0}\rangle+|e^{1}_{1,0}\rangle|e^{0}_{0,3}\rangle)+\frac{1}{2}|-,0\rangle(|e^{0}_{0,0}\rangle|e^{0}_{0,0}\rangle-|e^{1}_{1,0}\rangle|e^{0}_{0,3}\rangle)
\end{aligned}.
\end{equation}
If Eve wants to be undetected in this case, Alice's measurement results of these two qubits must be $|+\rangle$ and $|0\rangle$. That is, $U_F$ must satisfy the conditions:
\begin{equation}
|e^{0}_{0,0}\rangle=|e^{0}_{0,3}\rangle, \quad |e^{0}_{0,0}\rangle=|e^{1}_{1,0}\rangle.
\end{equation}

Putting everything together, if Eve's attack introduces no errors, it must satisfy:
\begin{equation}
\begin{aligned}
&\alpha=\mu_{0}=\mu_{3}=1,  \quad \beta=\nu_{0}=\nu_{3}=0, \\
&|e^{0}_{0,0}\rangle=|e^{0}_{0,3}\rangle=|e^{1}_{1,0}\rangle=|e^{1}_{1,3}\rangle.
\end{aligned}
\end{equation}

Now, we analyze the effect of $(U_E,U_F)$ on the qubits from the GHZ state $|G^{+}_{k_j}\rangle$. Assume Eve performs $(U_E,U_F)$ on the $i$-th qubit in $|G^{+}_{k_j}\rangle$, where the $i$-th qubit is denoted as $|a^{j}_{i}\rangle$ and $a^{j}_{i}\in \{0,1\}$. If Bob$_i$ performs the \textbf{Reflect} operation on the received qubit, after performing $(U_E,U_F)$, the state of $|a^{j}_{i}\rangle$ will evolve into
\begin{equation}
U_{F}U_{E}|a^{j}_{i}\rangle|0\rangle_E= 
\left\{
   \begin{aligned}
  & \mu_{0} |0,e^{0}_{0,0}\rangle +\nu_{0} |1,e^{1}_{0,0}\rangle \quad |a^{j}_{i}\rangle=\left|0\right\rangle \\
   &\nu_{3} |0,e^{0}_{1,3}\rangle +\mu_{3} |1,e^{1}_{1,3}\rangle \quad |a^{j}_{i}\rangle=\left|1\right\rangle  \\
   \end{aligned}
  \right..
\end{equation}
 If Bob$_i$ performs the \textbf{Measure-flip} operation on the received qubit, after performing $(U_E,U_F)$, the state of $|a^{j}_{i}\rangle$ will evolve into
\begin{equation}
U_{F}U_{E}|a^{j}_{i}\rangle|0\rangle_E= 
\left\{
   \begin{aligned}
  & \nu_{0} |0,e^{0}_{1,0}\rangle +\mu_{0} |1,e^{1}_{1,0}\rangle \quad |a^{j}_{i}\rangle=\left|0\right\rangle \\
   &\mu_{3} |0,e^{0}_{0,3}\rangle +\nu_{3} |1,e^{1}_{0,3}\rangle \quad |a^{j}_{i}\rangle=\left|1\right\rangle  \\
   \end{aligned}
  \right..
\end{equation}
Applying Eq. (23)  into Eqs. (24) and (25), we have
\begin{equation}
U_{F}U_{E}|a^{j}_{i}\rangle= |a^{j}_{i}\rangle |e^{0}_{0,0}\rangle.
\end{equation}
Thus, to avoid introducing errors, Eve's probe should be unassociated with the target qubit. That is to say, no matter what state the target qubit is in, Eve can only get the same result from her probe. Therefore, our protocol can withstand the entangle-measure attack.

\subsection{The Double-CNOT attack}
Semi-quantum protocols are susceptible to the Double-CNOT attack, which can be described as follows. Eve firstly intercepts the particle sent by Alice. Then, she performs the first attack $U_{CNOT}=|00\rangle \langle00|+|01\rangle \langle01|+|10\rangle \langle10|+|11\rangle \langle11|$, where the intercepted particle is the control bit and her ancillary particle $\left|0\right\rangle_E$ is the target bit. After that, Eve intercepts the particle sent by Bob$_i$ and executes the second $U_{CNOT}$ on the intercepted particle and her ancillary particle again. Finally, Eve measures her ancillary particle to obtain some useful information. 

However, Alice will detect Eve's attack in the eavesdropping checking since her attack will change the decoy photons' states. For the decoy photons, after performing the $U_{CNOT}$ operation, the qubit systems become:
\begin{equation}
U_{CNOT} \big(|0\rangle_A |0\rangle_E \big) =|00\rangle_{AE},
\end{equation}
\begin{equation}
U_{CNOT} \big(|1\rangle_A |0\rangle_E \big) =|11\rangle_{AE},
\end{equation}
\begin{equation}
U_{CNOT} \big(|+\rangle_A |0\rangle_E\big) =\frac{1}{\sqrt 2}\big(|00\rangle+|11\rangle \big)_{AE},
\end{equation}
\begin{equation}
U_{CNOT} \big(|-\rangle_A |0\rangle_E\big) =\frac{1}{\sqrt 2}\big(|00\rangle-|11\rangle \big)_{AE},
\end{equation}
where the subscripts $A$ and $E$ represent the Alice's qubit and Eve's ancillary particle, respectively. To eliminate the errors caused by the $U_{CNOT}$, Eve needs to perform the $U_{CNOT}$ again on the same control bit and the target bit. However, the qubit sent by Bob$_i$ is not the same as the qubit sent by Alice because Bob$_i$ performs the \textbf{Reorder} operation on the qubit sending to Alice. 
That is, Eve cannot eliminate the errors introduced by the $U_{CNOT}$ operation.
Thus, Eve's attack will inevitably introduce errors.

 For example, assume that the $j$-th qubit sent by Alice is $|+\rangle_A$ while the $j$-th qubit sent by Bob$_i$ is $|0\rangle_B$. After performing the Double-CNOT attack, the qubit systems are transformed as 
\begin{equation}
\begin{split}
U_{CNOT} &|0\rangle_B \big[U_{CNOT}(|+\rangle_A|0\rangle_E)\big]\\
\quad & =\frac{1}{\sqrt 2}\big( 
|000\rangle+|011\rangle \big)_{BAE}\\
& =\frac{1}{\sqrt 2} |0\rangle_B(|+\rangle|+\rangle+|-\rangle|-\rangle)_{AE}
\end{split}.
\end{equation}
Consider the scenario where Bob chooses the \textbf{Reflect} operation for $|+\rangle_A$. After undergoing the Double-CNOT attack, Alice will perform measurements on the particle reflected by Bob in the basis$\{|+\rangle,|-\rangle\}$. As indicated by Eq. (31), Alice's measurement results will randomly be in either $|+\rangle_A$ or $|-\rangle_A$, rather than being guaranteed to be in $|+\rangle_A$. During eavesdropping detection, Alice will discover this error. Therefore, Eve's attack will introduce errors and be detected by Alice with a non-zero probability.

 \subsection{The Trojan horse attack}
In our protocol, each qubit is transmitted twice. To extract Alice's and Bob's secret, Eve may perform Trojan horse attacks \cite{46,47,48} which mainly include the delay-photon attack and the invisible photon attack. Fortunately, by equipping with wavelength filters and photon number splitters \cite{49,50}, the Trojan horse attacks can be effectively defended. As a result, our protocol can withstand the Trojan horse attacks.

\subsection{The participant attack}

In SQSS protocols, dishonest ``classical'' users also pose a threat to the security of the protocol. They attempt to obtain the secret information acquired by other users through attack methods, thus independently reconstructing the shared secret information.
Here, we analyze an extreme case where $n-1$ dishonest participants conspire to steal Alice's secret without the honest participant's help. Assume that Bob$_t$ ($t=1,2,\dots,n$ and $t\neq i$) is dishonest and Bob$_i$ is honest. According to the steps of the protocol, Bob$_t$ will obtain part of the secret information, but this still cannot help them recover the secret information shared by Alice. To obtain the secret of Alice, the dishonest participants need to get the operations of Bob$_i$. Bob$_t$ may launch attacks on the transmitted qubits between Alice and Bob$_i$. However, in our protocol, there is not any qubit transmitted between Bob$_i$ and Bob$_t$. Obviously, Bob$_t$ is thoroughly independent of Bob$_i$. Thus, Bob$_t$ essentially acts as an external eavesdropper when she attacks the transmitted qubits between Alice and Bob$_i$. Based on the above analysis of Eve's attacks, Alice will detect Bob$_t$'s attacks with a non-zero probability. Therefore, our protocol can withstand  the participant attack.

\section{ Simulation and Discussion}
In this section, we will use IBM Qiskit to conduct circuit simulation of the proposed protocol and discuss the protocol performance. It is worth noting that, throughout the entire circuit simulation phase, we focus primarily on simulating the circuit corresponding to $(n+1)$-qubit GHZ state.
 
\subsection{Simulation based on IBM Qiskit}

The designed protocol is capable of enabling $n$ ``classical'' users to share secret information. For the sake of implementing simulations and demonstrating the correctness of the protocol, here, we use a three-party scenario (i.e., $n=3$) as an example to conduct circuit simulation of the protocol.

Firstly, let's introduce some basic settings for the three-party MSQSS protocol. Suppose there are three ``classical'' users, Bob$_1$, Bob$_2$, and Bob$_3$, and one quantum user, Alice. In this protocol, Alice intends to share the secret information $k=5$ with the ``classical'' users. The secret information can only be recovered when all ``classical'' users cooperate with each other.

According to the protocol's requirements, in the three-party MSQSS protocol scenario, Alice needs to prepare the $4$-qubit GHZ state $|G^{+}_{k=5}\rangle=\frac{1}{\sqrt 2}(|0101\rangle+|1010\rangle)$ as the transmission qubits. Following the steps of protocol, she keeps the first qubit while sending the remaining three qubits to Bob$_1$, Bob$_2$, and Bob$_3$, respectively.
Then, the classical users Bob$_1$, Bob$_2$, and Bob$_3$ will randomly perform \textbf{Measure-flip} or \textbf{Reflect} on the received particles and record the corresponding operation type. After that, Alice will measure the received particles in basis $\{|0\rangle, |1\rangle\}$.

Below, we simulate the circuit for the aforementioned process. Here, we consider three scenarios: one where all "classical" users perform a \textbf{Measure-flip} operation on the received particles, another where they execute \textbf{Reflect} operations on all received particles, and the last one that both \textbf{Measure-flip} and \textbf{Reflect} operations are performed on the received particles.
\begin{figure*}[h]
  \centering
  \begin{subfigure}{0.6\textwidth}
    \centering
    \includegraphics[width=\linewidth, height=1.8in]{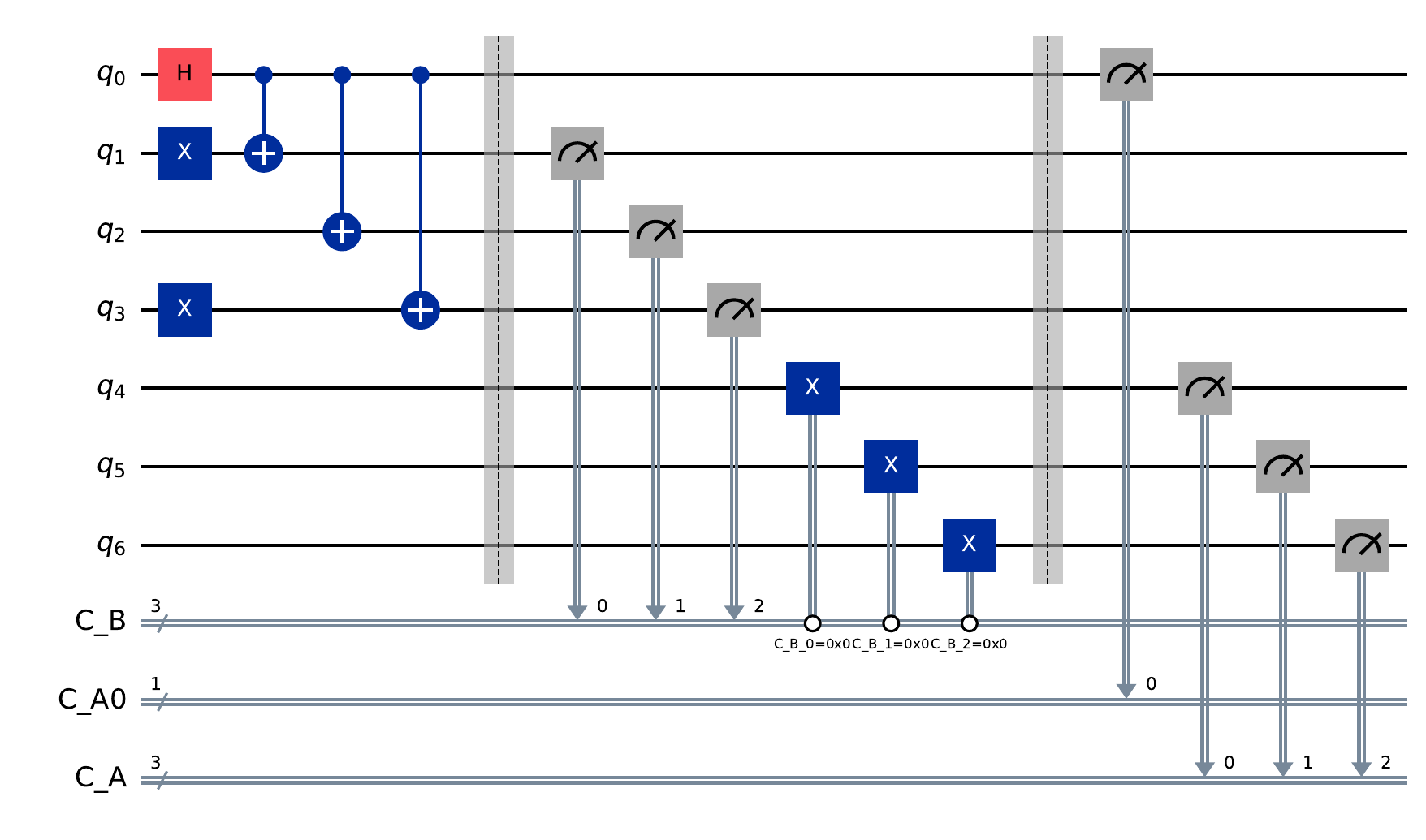}
   \caption{ }
  \end{subfigure}%
  \hfill
  \begin{subfigure}{0.35\textwidth}
    \centering
    \includegraphics[width=\linewidth, height=1.5in]{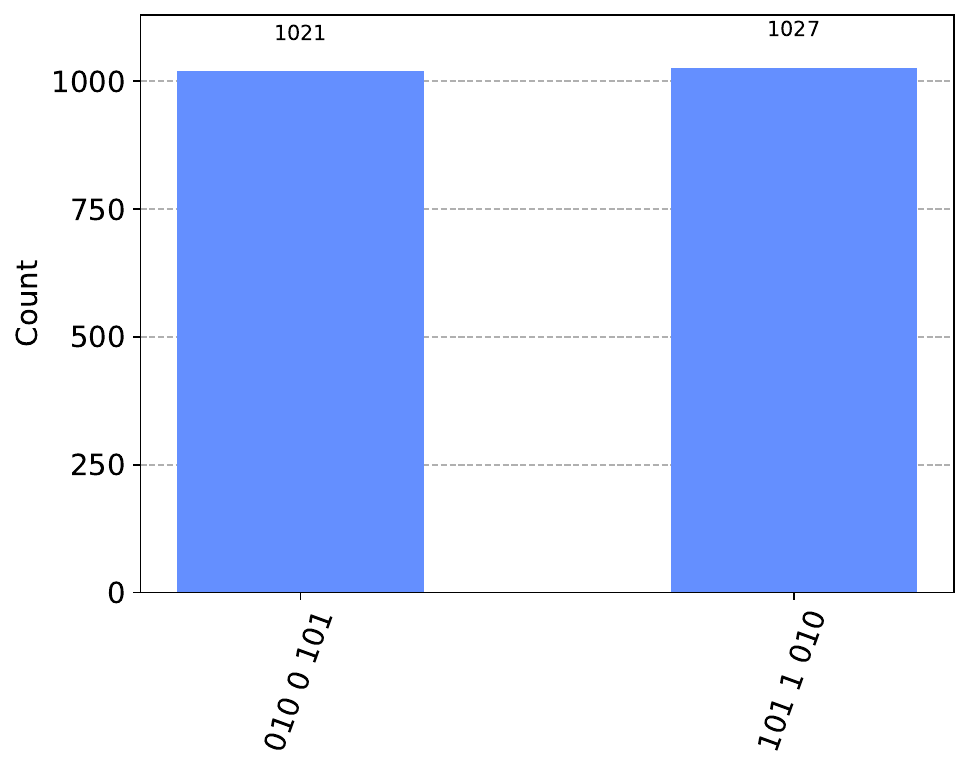}
    \caption{ }
  \end{subfigure}
  \caption{Quantum circuit diagram corresponding to all ``classical'' users performing measure-flip operations. Registers $(q_0,q_1,q_2,q_3)$ are used to generate 4-qubits GHZ state $|G^{+}_{k=5}\rangle$. Registers $(q_4,q_5,q_6)$ are used by ``classical'' users to prepare new quantum states after performing measure-flip operations. Classical register C\_B records the measurement results of ``classical'' users on the received particles, C\_A0 records the measurement results of particles retained by Alice, and C\_A records the measurement results of particles after being operated on by ``classical'' users.   }
  \end{figure*}
  
Fig. 1 shows the corresponding quantum circuit and simulation results when all ``classical'' users perform \textbf{Measure-flip} operations. In Fig. 1(a), the section before the first barrier represents Alice preparing the quantum state $|G^{+}_{k=5}\rangle$, the middle section illustrates ``classical'' users preparing particles in the opposite state after measuring the received particles, and the final part represents Alice performing single-particle measurements on the received particles. 
The simulation results indicate that the designed circuit meets the requirements of the protocol. Taking the first column of Fig. 1(b) as an example, the measurement result is "010 0 101", which represents Alice's measurement result of the measure-flipped particles as "010", the measurement result of the particle retained in her hand as "0", and the measurement results of Bob$_1$, Bob$_2$, and Bob$_3$ as "101". Because Bob$_1$, Bob$_2$, and Bob$_3$ all performed \textbf{Measure-flip} operations, thus Alice's measurement results should be opposite to Bob's measurement results, and the simulation results also confirm this. Additionally, the measurement results of the particles retained in Alice's hand and received by Bob$_1$, Bob$_2$, and Bob$_3$ as "0101" are consistent with the measurement results of $|G^{+}_{k=5}\rangle$.

In order to be consistent with the protocol description, in this case, Alice's measurement results of the received particles and the particle kept in her own hands can be recorded with the symbol $m_1=0, m_2=1,m_3=0,m_0=0 $, respectively. The operations corresponding to Bob$_1$, Bob$_2$, and Bob$_3$ are represented by the symbol $r^\prime_1=1,r^\prime_2=1,r^\prime_3=1$ respectively. Then, according to protocol step 5, Alice will use the value of $m_0$ to perform the XOR operation on $m_1$, $m_2$ and $m_3$ to obtain $m^\prime_1=m_0\oplus m_1=0$, $m^\prime_2=m_0\oplus m_2=1$ and $m^\prime_3=m_0\oplus m_3=0$ and publish them to the ``classical'' user.
Finally, Bob$_1$, Bob$_2$, and Bob$_3$ can recover Alice's secret information $k=5$ as follows:
\begin{equation}
\begin{aligned}
k&=(m^{\prime }_{1}\oplus r^{\prime }_{1})\times2^{2}+(m^{\prime}_{2}\oplus r^{\prime }_{2})\times2+\dots+(m^{\prime }_{3}\oplus r^{\prime }_{3})\times2^{0}\\
&=(0\oplus 1)\times2^{2}+(1\oplus 1)\times2+\dots+(0\oplus 1)\times2^{0}
\\
&=5
\end{aligned}.
\end{equation}
\begin{figure*}[h]
  \centering
  \begin{subfigure}{0.6\textwidth}
    \centering
    \includegraphics[width=\linewidth, height=1.7in]{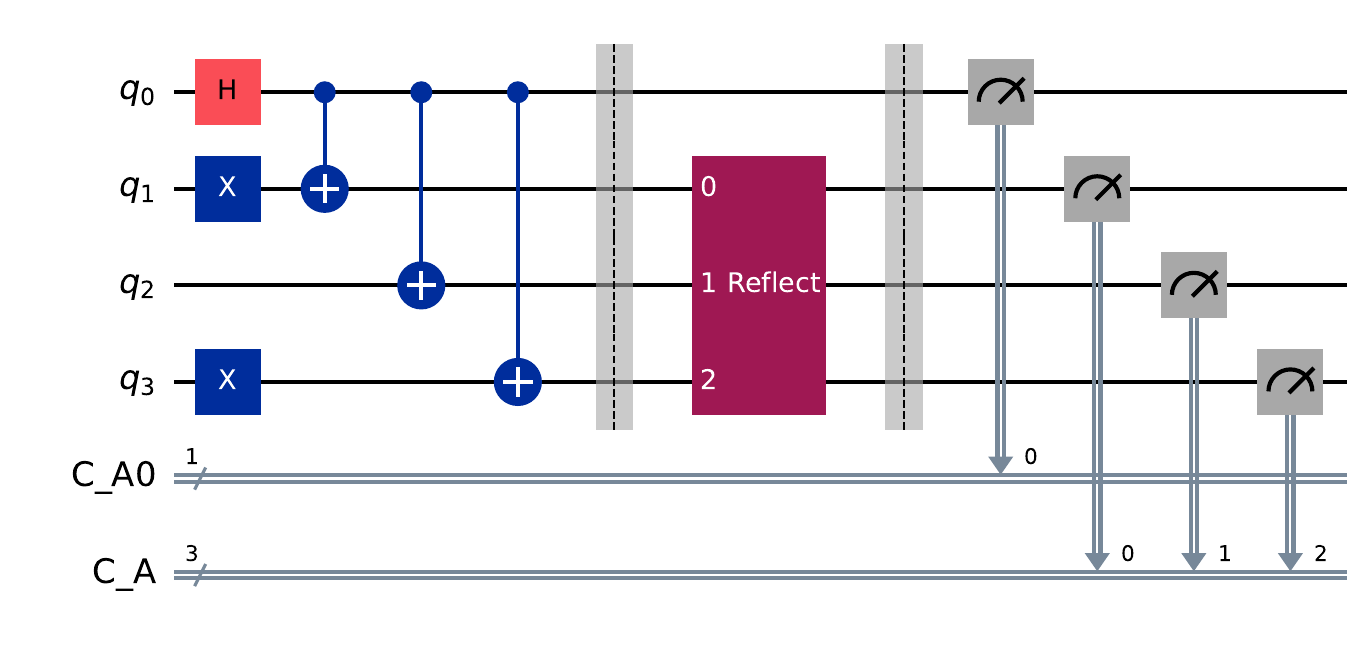}
   \caption{ }
  \end{subfigure}%
  \hfill
  \begin{subfigure}{0.35\textwidth}
    \centering
    \includegraphics[width=\linewidth, height=1.5in]{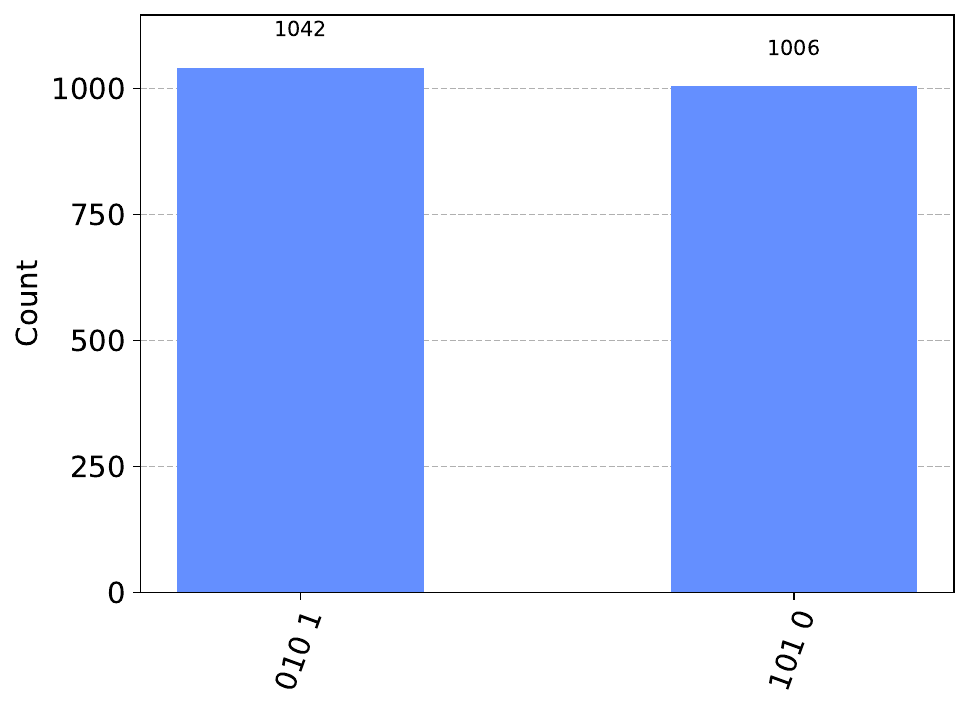}
    \caption{ }
  \end{subfigure}
  \caption{Quantum circuit diagram corresponding to all ``classical'' users performing reflect operations.   }
  \end{figure*}
  
Then, we consider the case that ``classical'' users perform \textbf{Reflect} operations on all received particles. Fig. 2 shows the corresponding quantum circuit and simulation results of this case.  In this scenario, Alice essentially performs single-particle measurements on the 4-qubit GHZ state $|G^{+}_{k=5}\rangle$. Assuming Alice obtains the measurement result of "1" for the particle she retains, then the other three particles' results will be "010". This outcome is consistent with the first column of the simulation results shown in Fig. 2(b). Therefore, it indicates that the designed circuit meets the requirements of the protocol. Let $m_0=1,m_1=0$, $m_2=1$ and $m_3=0$ represent Alice's measurement results. Then, after XOR operation, we can get $m^\prime_1=m_0\oplus m_1=1$, $m^\prime_2=m_0\oplus m_2=0$ and $m^\prime_3=m_0\oplus m_3=1$. Moreover, in this case, Bob$_1$, Bob$_2$, and Bob$_3$ perform \textbf{Reflect} operations, thus leading to the conclusion that $r^\prime_1=0,r^\prime_2=0,r^\prime_3=0$. Integrating all the information, the ``classical'' users can calculate Alice's secret information as follows:
\begin{equation}
\begin{aligned}
k&=(m^{\prime }_{1}\oplus r^{\prime }_{1})\times2^{2}+(m^{\prime}_{2}\oplus r^{\prime }_{2})\times2+\dots+(m^{\prime }_{3}\oplus r^{\prime }_{3})\times2^{0}\\
&=(1\oplus 0)\times2^{2}+(0\oplus 0)\times2+\dots+(1\oplus 0)\times2^{0}
\\
&=5
\end{aligned}.
\end{equation}

Furthermore, we consider a more general scenario where "classical" users perform random \textbf{Measure-flip} or \textbf{Reflect} operations on the received particles. Here, we assume that Bob$_1$, Bob$_2$, and Bob$_3$ respectively apply \textbf{Measure-flip}, \textbf{Reflect}, and \textbf{Reflect} operations to the received particles. The circuit simulation for scenarios corresponding to different operations is similar. Figure 3 illustrates the corresponding quantum circuit and simulation results for this scenario. Similarly, we assume that Alice retains the measurement result of 0 on her own particles, i.e., the result recorded in register C\_A0 is 0, corresponding to the first column in Figure 3b. From the simulated results in Figure 3, it can be observed that the measurement results of the particles after Bob$_1$, Bob$_2$, and Bob$_3$'s operations are 001, indicating consistency with the operations performed. Let $m_0=0,m_1=0$, $m_2=0$ and $m_3=1$ represent Alice's measurement results. Then, after XOR operation, we can get $m^\prime_1=m_0\oplus m_1=0$, $m^\prime_2=m_0\oplus m_2=0$ and $m^\prime_3=m_0\oplus m_3=1$. Moreover, following the ``classical'' users' operation, we can obtain $r^\prime_1=1,r^\prime_2=0,r^\prime_3=0$. Integrating all the information, the ``classical'' users can calculate Alice's secret information as follows:
\begin{equation}
\begin{aligned}
k&=(m^{\prime }_{1}\oplus r^{\prime }_{1})\times2^{2}+(m^{\prime}_{2}\oplus r^{\prime }_{2})\times2+\dots+(m^{\prime }_{3}\oplus r^{\prime }_{3})\times2^{0}\\
&=(0\oplus 1)\times2^{2}+(0\oplus 0)\times2+\dots+(1\oplus 0)\times2^{0}
\\
&=5
\end{aligned}.
\end{equation}

\begin{figure*}[h]
  \centering
  \begin{subfigure}{0.6\textwidth}
    \centering
    \includegraphics[width=\linewidth, height=1.7in]{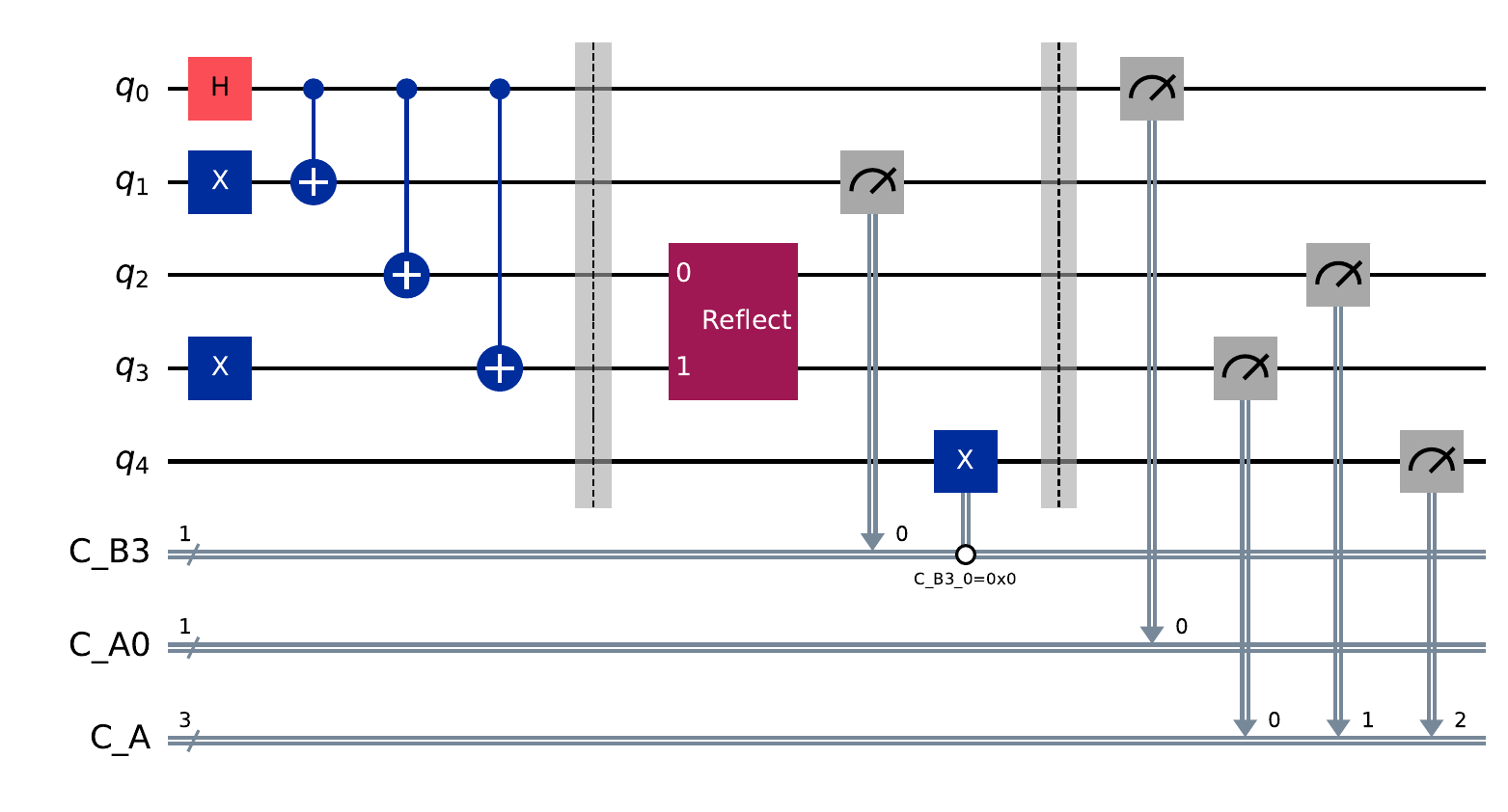}
   \caption{ }
  \end{subfigure}%
  \hfill
  \begin{subfigure}{0.35\textwidth}
    \centering
    \includegraphics[width=\linewidth, height=1.5in]{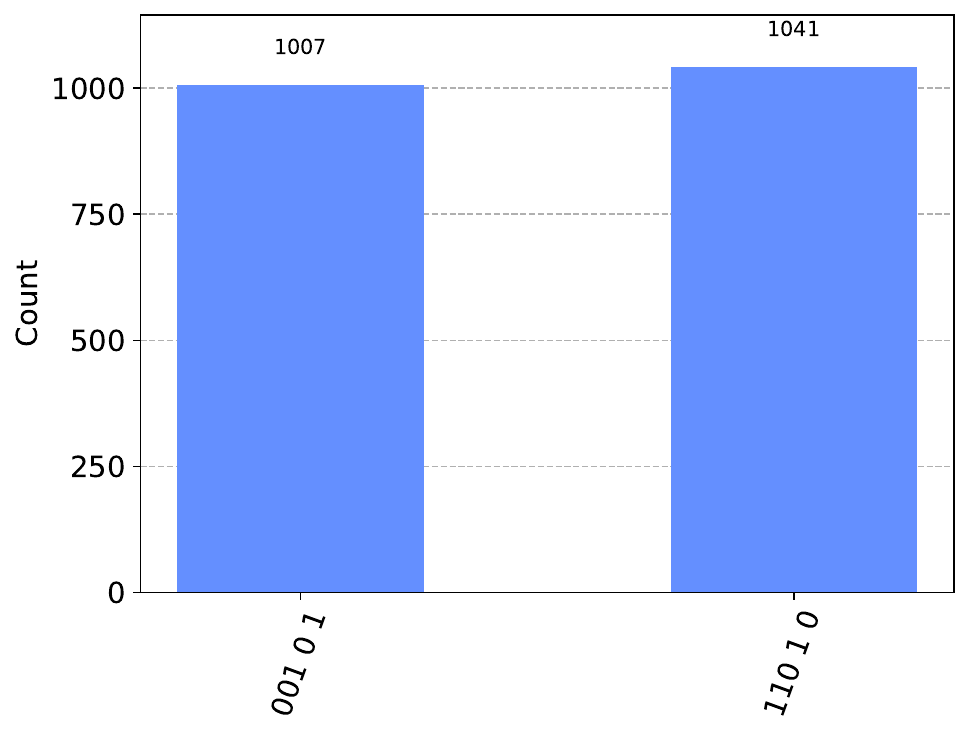}
    \caption{ }
  \end{subfigure}
  \caption{Quantum circuit diagram corresponding "classical" user performs measure-flip and reflect operations.  }
  \end{figure*}

Above, we performed circuit simulations on three situations where ``classical'' users perform \textbf{Measure-flip} and \textbf{Reflect} operations on the receiving particles, and verified the correctness of the protocol. The results show that the designed protocol is feasible and correct.

\subsection{Discussion}
This part will evaluate the performance of the protocol in terms of scalability, qubit efficiency, and shared secret information types.

Firstly, the proposed MSQSS protocol can meet the requirement of secret sharing among any number of "classical" users, which is a significant advantage compared to traditional two-party SQSS protocols\cite{31,32,33,34,35,36} that can only accommodate two "classical" users, thus having certain limitations. Additionally, the designed protocol exhibits scalability. It can dynamically adjust according to actual needs to accommodate information sharing among different numbers of users. Therefore, whether it is for small-scale or large-scale secret information sharing requirements, the protocol can flexibly adapt and provide reliable solutions for users.

Qubit efficiency is an important metric for evaluating protocol performance. For the sake of uniform comparison, we focus on the qubit efficiency comparison between multi-party SQSS protocols\cite{37,38,39}. Here, qubit efficiency is defined as $\eta=c/q$, where $c$ denotes the shared classical bits, and $q$ denotes the transmitted qubits. In our protocol, Alice wants to share $L$ classical secret messages to $n$ classical users. That is, $c=L$. For the transmitted qubits, Alice needs to generate $L$ $(n+1)$-qubit GHZ states and $L\times n$ decoy photons. While $n$ classical users need to generate $L$ qubits respectively to replace the qubits measured by them. That is, $q=L(n+1)+2Ln$. Thus, the qubit efficiency of the proposed protocol is $\frac{1}{3n+1}$. Similarly, the qubit efficiency of Refs. \cite{37,38,39} can be calculated, and the specific results are shown in Table 2. 

\begin{table}
\centering
\label{t2}  
\caption{The comparisons between MSQSS protocols}
\begin{tabular}{p{0.2\linewidth}<{\centering}p{0.1\linewidth}<{\centering}p{0.18\linewidth}<{\centering}p{0.1\linewidth}<{\centering}p{0.16\linewidth}<{\centering}} 
\Xhline{1.0pt}\noalign{\smallskip}
  & Ref.\cite{37}  & Ref.\cite{38}  &   Ref.\cite{39}  &  Our protocol\\
\noalign{\smallskip}\hline\noalign{\smallskip}
 Quantum resource  &  Bell states  &  $(n+1)$-qubit GHZ-like states &  Bell states & $(n+1)$-qubit GHZ states\\
 \noalign{\smallskip}
 Number of classical users   & $n$  & $n$   &  $n$ & $n$ \\
  \noalign{\smallskip}
 Single-particle measurement  &  Yes & Yes & Yes &  Yes\\
 \noalign{\smallskip}  
 Entangled state measurement  &  Yes & Yes &  Yes &  No\\
 \noalign{\smallskip}  
 The reflected particles  as the secret key  &  No   & No &  No & Yes\\
 \noalign{\smallskip}
  Share specific message  &  No   & No &  No & Yes\\
 \noalign{\smallskip}
 Qubit efficiency   & $\frac{1}{2^{n+1}+n2^{n-1}}$  & $\frac{1}{3n+2}$   &  $\frac{1}{5n}$  & $\frac{1}{3n+1}$\\
\noalign{\smallskip}
\Xhline{1.0pt}
\end{tabular}
\end{table}

In terms of the type of shared secret information,  our protocol offers a distinct advantage that it can share specific information among $n$ classical users, whereas the previous MSQSS protocol was limited to sharing random secret information. Sharing specific secret information is necessary in some secure multi-party scenarios, such as in the financial field or the healthcare field, where transaction information and patient privacy data shared by multiple parties need to be deterministic. Our protocol can meet the task requirements of these specific scenarios and provides users with a reliable solution for secure multi-party collaboration.

In addition,  in our protocol, the secret keys are constructed by the classical users' operations. That is, both the measured qubits and reflected qubits can be used as the secret keys. While in previous MSQSS protocols, the reflected qubits are usually used for eavesdropping checks, not as secret keys. By increasing the utilization of transmission particles, the efficiency of the protocol is improved. Table 2 clearly shows that our protocol has higher qubit efficiency.

\section{Conclusion }
In this paper, by utilizing the $(n+1)$-qubit GHZ states and the idea of semi-quantum, a multi-party semi-quantum secret sharing protocol is proposed, where the specific messages can be shared by quantum Alice to $n$ ``classical'' parties (Bob$_i$, $i=1,2,\dots,n$). Only the ``classical'' parties cooperate together can recover Alice's messages. Unlike previous protocols that rely on measurement outcomes as keys, this protocol establishes keys among users by utilizing measure-flip and reflect operations performed by ``classical'' users. This effectively improves the utilization of transmitted particles, providing a new approach to enhance the qubit efficiency of multi-party semi-quantum protocols.  Furthermore, our protocol is feasible with existing technologies since complex entanglement measurements and unitary operations are not required. The relevant circuit simulation also shows that the designed protocol is correct and feasible. The proposed protocol is resistant to most typical external and internal attacks and has advantages in terms of scalability, qubit efficiency and type of shared message. Therefore, this protocol may be more consistent with the needs of practical secret sharing tasks.

In future research, we will explore the application scenarios and performance optimization of semi-quantum secret sharing protocols. In practical secret sharing scenarios, the addition and removal of users are common occurrences. However, current research can only achieve secret sharing between fixed users. Therefore, how to dynamically realize semi-quantum secret sharing will be the focus of the next stage.
\\
\\
\\
\textbf{Acknowledgments} 
This work was supported in part by the Key Research and Development Program of Ningxia Hui Autonomous Region, grant number ``2021BEG02007'' and the Open Research Fund of Key Laboratory of Cryptography of Zhejiang Province, grant number ``No. ZCL21006''.

\end{document}